\documentclass{article}
\usepackage{spconf,amsmath,graphicx}
\usepackage{tikz}

\usepackage{comment}
\usepackage{amssymb}
\usepackage{latexsym}
\usepackage{url}
 \usepackage[nodots]{numcompress}

\usepackage{amsfonts}         
\usepackage{amsmath}
\usepackage{amssymb}
\usepackage{amsthm}
\usepackage{bm}
\usepackage{caption}
\usepackage{subcaption}
\usepackage{booktabs}         
\usepackage{nicefrac}       	
\usepackage{microtype}      
\usepackage{centernot}		
\usepackage{threeparttable}
\usepackage{enumitem, hyperref, cleveref}
\setlist[description]{leftmargin=7pt}
\hypersetup{pdfborder=0 0 0,colorlinks=true}
\crefname{equation}{eqn.}{eqns.}
\crefname{figure}{fig.}{figs.}
\crefname{table}{tab.}{tabs.}


\makeatletter
\def\namedlabel#1#2{\begingroup
    #2%
    \def\@currentlabel{#2}%
    \phantomsection\label{#1}\endgroup
}
\makeatother

\usepackage{enumitem}
\setlist[enumerate]{leftmargin=.5in}
\setlist[itemize]{leftmargin=.5in}

\newcommand{\ie}{i.e., }
\newcommand{\eg}{e.g., }

\usepackage{amsopn}
\graphicspath{{figures/}}
\newcommand {\R}   {{\rm I\!R}}

\newcommand{\ds}{\displaystyle}

\newcommand{\Expt}[1]{\mathbb{E}\!\left[{#1}\right]}

\newcommand{\bfK}{{\bf K}}

\renewcommand{\bfK}{{\bf K}}
\newcommand{\bfA}{{\bf A}}

\newcommand{\bfI}{{\bf I}}

\newcommand{\bfX}{{\bf X}}

\newcommand{\bfm}{{\bf m}}
\newcommand{\bfr}{{\bf r}}

\newcommand{\bff}{{\bf f}}

\newcommand{\bfy}{{\bf  y}}
\newcommand{\bfx}{{\bf  x}}

\newcommand{\bfepsilon}{\boldsymbol \epsilon}

\usepackage{mathtools}
\usepackage{url}

\newcommand{\bfSigma}{{\boldsymbol \Sigma}}

\usepackage{array}

\newlength{\myrowheight}
\newcommand{\kse}{k_{\text{SE}}}
\newcommand{\kmk}{k_{\text{MK}}}

\usepackage{color}

  \newcommand{\doneone}[2][]{}

\newcommand{\removed}[1]{}
\newcommand{\done}[2][]{}
\newcommand{\mdone}[1]{}

\title{Simultaneous Reconstruction and Uncertainty Quantification for Tomography}
\name{Agnimitra Dasgupta\textsuperscript{$\star$}, Carlo Graziani\textsuperscript{$\dagger$}, and Zichao Wendy Di\textsuperscript{$\dagger$}\thanks{
		 This material was based upon work supported by the U.S. Department of Energy, Office of Science, Office of Advanced Scientific Computing Research, Scientific Discovery through Advanced Computing (SciDAC) program through the FASTMath Institute under Contract DE-AC02-06CH11357 at Argonne National Laboratory.
}}
\address{\textsuperscript{$\star$}Sonny Astani Department of Civil \& Environmental Engineering,
	University of Southern California. \\ \textsuperscript{$\dagger$}Mathematics and Computer Science Division, Argonne National Laboratory.}
\begin{document}
	%
	\maketitle
	\begin{abstract}
		Tomographic reconstruction, despite its revolutionary impact on a wide range of applications, suffers from its ill-posed nature in that there is no unique solution because of limited and noisy measurements. Therefore, in the absence of ground truth, quantifying the solution quality is highly desirable but under-explored. In this work, we address this challenge through Gaussian process modeling to flexibly and explicitly incorporate prior knowledge of sample features and experimental noises through the choices of the kernels and noise models. Our proposed method yields not only comparable reconstruction to existing practical reconstruction methods (\eg regularized iterative solver for inverse problem) but also an efficient way of quantifying solution uncertainties. We demonstrate the capabilities of the proposed approach on various images and show its unique capability of uncertainty quantification in the presence of  various noises. 
	\end{abstract}
	\section{Introduction}
	\label{sec:intro}
	\vspace{-0.5em}
	Tomographic imaging refers to the reconstruction of a 3D object from its 2D projections by sectioning the object, through the use of any kind of penetrating wave, from many different directions. It has shown revolutionary impacts in a number of fields ranging from biology, physics, and chemistry to astronomy \cite{maire2014quantitative,sharif2020comprehensive}.  The technique requires an accurate image reconstruction, however, the resulting reconstruction problem is ill-posed due to insufficient and noisy measurements \cite{di2017joint}, in that there is no unique reconstruction. Therefore, quantifying the solution quality is critical but challenging. Traditionally, ill-posedness is tackled using regularization techniques that incorporate prior knowledge. However, regularization choices including the type of regularizer and its corresponding weight can be difficult and lead to substantial variations in reconstruction, especially in the absence of ground truth \cite{antil2020bilevel}.
	
	Instead, we consider tomographic reconstruction explicitly from a Bayesian statistical perspective, utilizing \textit{Gaussian process} (GP) modeling \cite{williams2006gaussian} to simultaneously achieve high-quality reconstruction and uncertainty quantification. In such a setting, ill-posedness is rephrased as uncertainty and qualitatively quantified by the posterior variance. GP-based reconstruction establishes a prior model over the unknown object that is comprised of a \text{mean} and \text{covariance} function. The GP prior is controlled by hyperparameters that take over the role of regularization parameters, some of which represent multiple length scales embedded in the object. The predictions from the model can be compared with observational data by means of the \textit{likelihood}, where noise can be naturally incorporated. We determine the model's hyperparameters by maximizing the likelihood, analogous to a data fit approach to regularization. The resulting reconstruction is obtained in the form of a posterior mean, attended by an estimated uncertainty encoded in the posterior covariance. In particular, we examine the effect of heteroscedastic (i.e., non-i.i.d) measurement noise and the advantage of modeling such noise correctly. We also explore the use of kernel mixtures to flexibly model any length-scale mixing while avoiding parameter proliferation. For simplicity, we confine ourselves to reconstructing 2D images. Extension to the 3D case is possible by considering individual slices.
	
	\noindent \textbf{Related Work:} GP modeling has been explored in various imaging problems such as learning-based image super-resolution \cite{he2011single,wang2015single}. Svensson et al.~\cite{li2013bayesian} applied GP to soft x-ray tomography to reconstruct emissivity in stellarator plasmas, with assumed homoscedastic (\ie i.i.d.) measurement noise. To capture length scale variations, non-stationary GP kernels were used. The resulting model with a large number of parameters prevents its application being practical. ``GP-like'' priors have been explored to inform the choice of the regularizer for tomographic reconstruction due to the close connection between regularization and prior distribution \cite{venkatakrishnan2016robust}. Purisha et al.~\cite{purisha2019probabilistic} demonstrated GP reconstruction in a continuous tomographic setting without practical assumptions such as heteroscedastic measurement noise. Their approach uses Markov chain Monte Carlo to account for uncertainty in the hyperparameters of the GP prior, and requires a careful selection of priors for these hyperparameters. Therefore, their approach can quickly become impractical especially in the case of heteroscedastic measurement noise due to the very large degrees of freedom and often unavailable prior knowledge of the hyperparameters. 
	
	\section{Method}
	\vspace{-0.5em}
	Consider the physical domain $\Omega \subseteq \R^2 $ and the compactly
	supported function $f : \Omega \to \mathbb{R}$ representing the physical
	property of the object one is interested in recovering (e.g., the elemental density in emission imaging). In practice, we cannot recover the desired object property at all points in space. Instead, we discretize $\Omega$ (containing the compact object) into $n \times n$ pixels, where the set of coordinates of each pixel center is $\bfX = \{\bfx_i\}_{i=1}^{n^2} \in \R^{n^2}$, $\bfx_i \in \Omega$. The vectorized object to be recovered is denoted as $\bff = \{\bff_i\}_{i=1}^{n^2}\in \R^{n^2}$, where $\bff_i = f(\bfx_i)$. We impose a GP prior on $f(\cdot)$ by introducing a mean function $m(\cdot)$ and a covariance function $k(\cdot,\cdot)$. These functions are
	parameterized by external hyperparameters controlling features such as length scales, range of variability, and smoothness,  as we will show below. Therefore, the GP model gives rise to a multivariate normal prior distribution on $\bff$, $\bff\sim\mathcal{N}(\bfm,\bfK)$, with prior mean vector $\bfm \in \R^{n^2}$ and the prior covariance matrix $\bfK \in \R^{n^2 \times n^2}$. 
	
	For the experimental configuration, we index a beam by  $\tau$ and an angle by $\theta$. During the experiment, the object is rotated by $N_\theta$ angles, where at each angle it is raster scanned by $N_\tau$ beams \cite{di2017joint}. We denote the vectorized set of measurements (i.e., the ``sinogram'')  by $\ds \bfy=[\bfy_t]_{t=1}^{N_\theta N_\tau}$. The forward model is then governed by the discrete Radon transform \cite{beylkin1987discrete}, as follows: 
	$\bfy=\bfA\bff + \bfepsilon$ ,
	where $\bfA=[\bfA_{t,i}]\in \R^{N_\theta N_\tau \times n^2}$ and $\bfA_{t,i}$ is the intersection length\footnote{In principle, the components of $\bfA$ are proportional to the intersection length, with the proportionality constant determined by beam intensity and detector efficiency. For the sake of simplicity, and without loss of generality, we set this constant to 1 here.} of the $t =\theta+(\tau-1)N_\theta{\,\textsuperscript{th}}$ beam with the
	$i$\textsuperscript{th} pixel. The measurement error $\bfepsilon$ is assumed to be a zero-mean multivariate normal random variable with covariance $\Expt{\bfepsilon\bfepsilon^T}=\bfSigma_\epsilon$.\footnote{We use $\Expt{\cdot}$ to denote the expected value of a random variable.}  In general, $\bfepsilon$ is assumed to be statistically independent
	of the random variable $\bff$, but it can be heteroscedastic or even correlated (i.e., $\bfSigma_\epsilon$ may have nonzero off-diagonal entries). We consider only the uncorrelated case in this work, with a particular focus on heteroscedastic noise (i.e.,  the diagonal elements of $\bfSigma_\epsilon$ may differ from each other). Such noise is critical and of practical interest in experimental situations, when (for example) measurement errors arise in the Gaussian limit of Poisson photon counting statistics.
	
	Since $\bff$ is a normal random variable and $\bfy$ is linearly dependent on $\bff$, the pair $[\bfy,\bff]$ is necessarily governed by a
	normal distribution. Noting that $\Expt{\bff}=\bfm$,
	$\Expt{\bfy}=\bfA\Expt{\bff}=\bfA\bfm$,
	$\Expt{(\bff-\bfm)(\bff-\bfm)^T}=\bfK$, and given the statistical
	independence of $\ds \bff$ and $\bfepsilon$ (i.e., $\Expt{\bff\bf\epsilon^T}=0$), we
	can obtain \\
	$\Expt{(\bfy-\bfA\bfm)(\bff-\bfm)^T}=\bfA\Expt{(\bff-\bfm)(\bff-\bfm)^T}=\bfA\bfK$,
	and $\Expt{(\bfy-\bfA\bfm)(\bfy-\bfA\bfm)^T}=\bfA\bfK\bfA^T+\bfSigma_\epsilon$. 
	Then, the resulting joint normal distribution is
	\begin{equation}
		\begin{Bmatrix}
			\mathbf{y} \\
			\mathbf{f}
		\end{Bmatrix} \sim \mathcal{N}
		\begin{pmatrix}
			\begin{Bmatrix}
				\mathbf{Am}\\
				\mathbf{m}
			\end{Bmatrix},
			\begin{bmatrix}
				\mathbf{AKA}^{\mathrm{T}} + \mathbf{\Sigma}_{\epsilon} & \mathbf{AK} \\
				\mathbf{KA}^{\mathrm{T}} & \mathbf{K}
			\end{bmatrix}
		\end{pmatrix}.
		\label{eqn:jointDis}
	\end{equation}
	
	A simple choice for the GP prior mean function is a constant function; \ie, $m(\mathbf{x}) = c,\forall\, \bfx \in \Omega$. Then, $\mathbf{m} = c\mathbf{1}$, where $\mathbf{1}$ is a vector of all ones and $c$ is a hyperparameter. An observation of the sinogram $\bfy$ then yields the posterior predictive distribution \cite{williams2006gaussian} as
	$
	\mathbf{f}|\mathbf{y} \sim \mathcal{N}(\bfm^*, \bfK^*),
	$
	where
	
	\begin{align}
		\label{post_1}
		&\bfm^* = \mathbf{m} + \mathbf{KA}^{\mathrm{T}}[ \mathbf{AKA}^{\mathrm{T}} + \mathbf{\Sigma}_{\epsilon} ]^{-1} (\mathbf{y} - \mathbf{Am}),\\
		&\bfK^* = \mathbf{K} - \mathbf{KA}^{\mathrm{T}}[ \mathbf{AKA}^{\mathrm{T}} + \mathbf{\Sigma}_{\epsilon} ]^{-1} \mathbf{AK}.
	\end{align}
	are the posterior mean vector and covariance matrix, respectively. The diagonal elements of $\bfK^*$ provides the posterior variance in the reconstruction of each object pixel and a larger magnitude indicates greater uncertainty.
	
	The choice of covariance function (i.e., kernel) $k(\mathbf{x}, \mathbf{x}')$ encodes assumptions about object function $f$ such as smoothness and differentiability, and is critical for robust performance. Popular choices include stationary covariance functions, which depend only on the metric distance $r$ between any two points  $\bfx, \bfx' \in \Omega$ \cite[Chapter 4]{williams2006gaussian}. Given $\sigma_f$ as the prior variance at each location, we focus on two popular stationary kernels including the squared exponential (SE)
	kernel 
	$\kse(r; l) = \sigma_f^2 \exp\left(-r^2/2\right)  
	$ and the Mat\'ern kernels (MK) 
	$\kmk(r; l) = \sigma_f^2\frac{2^{1 -\nu}}{\Gamma(\nu)} \big( \sqrt{2\nu}r\big)^\nu B_\nu \big( \sqrt{2\nu}r \big),$
	where $\nu>0$ is a constant and $B_\nu$ is the modified Bessel function
	of the second kind. The SE kernel $\kse$ is appropriate for describing infinitely differentiable functions, as suggested by the fact that $\ds \lim_{\nu \to \infty}\kmk = \kse$ \cite{williams2006gaussian}. For the MK kernel $\kmk$, $\nu$ controls the level of smoothness of $f$, since the order of differentiability of functions sampled from
	a GP with a Mat\'ern kernel is $\lfloor\nu\rfloor$.  We particularly focus on two common types of the MK kernels: MK32 (i.e., $\nu = 3/2$) and MK52 (i.e., $\nu = 5/2$).
	
	While the GP methodology offers the flexibility to use any distance metric \cite{ba2012composite}, we choose the isotropic Euclidean distance $\ds r = \sqrt{(\mathbf{x} - \mathbf{x'})^{\mathrm{T}}(l^2\mathbf{I})^{-1}(\mathbf{x} - \mathbf{x'})}, \, \forall \, \bfx, \bfx' \in \Omega$, where $l$ represents the length scale along coordinate axes. To further capture variations on multiple length scales present in many real objects, we consider composite kernels obtained by convex superposition of two or more composite kernels $k_i$, 
	$ k(r) = \sum_{i=1}^{N_k} \sigma_{f,i}^2 k_{i}(r; l_i).$ 
	For notational compactness, we denote by $\bm{\sigma}_f$ and $\bm{\ell}$ the $N_k$-dimensional vectors with components $\sigma_{f,i}$ and $l_i$, respectively. We particularly focus on homogeneous composite kernels, that is, with all component kernels of the same type.
	
	The parameters $\bm{\sigma}_f$ and $\bm{\ell}$, associated with kernel $k$,  along with the prior constant mean $c$ are together referred to as the hyperparameters. The choice of hyperparameters is critical for the performance of the GP method and must be chosen judiciously. Popular parameter-tuning strategies such as
	cross-validation \cite{rippa1999algorithm, montegranario2014radial} require extensive trial and error and can be computationally inefficient. Instead, we optimize hyperparameters by maximizing the marginal likelihood of $\bfy$ \cite{williams2006gaussian}. To be more specific,
	given $\bfy\sim\mathcal{N}(\bfA\bfm,\bfA\bfK\bfA^T+\bfSigma_\epsilon)$ (see \Cref{eqn:jointDis}),  the negative log-likelihood function of $\bfy$ is 
	\begin{equation*} 
		\ds \mathcal{J}(c,\bm{\sigma}_f,\bm{\ell}) =\ds \frac{1}{2}\mathbf{r}^{\mathrm{T}}\mathbf{K}_y^{-1}\mathbf{r} + \frac{1}{2}\log|\bfK_y| + \frac{m}{2}\log 2\pi ,
	\end{equation*}
	where $\mathbf{r} = \mathbf{y} - \mathbf{Am}$ is the residual and $\bfK_y=\bfA
	\bfK \bfA^T + \mathbf{\Sigma}_{\epsilon}$. The optimal parameters $[c^*,\bm{\sigma}_f^*,\bm{\ell}^*]$ are
	obtained by minimizing the following constrained optimization problem:
	\begin{equation}
		\label{optim1}
		[c^*,\bm{\sigma}_f^*,\bm{\ell}^*] = \operatorname*{arg\,min}_{c,\bm{\sigma_f},\bm{l}} \mathcal{J}(c,\bm{\sigma}_f,\bm{\ell}) \, \text{s.t.}  \, \sigma_{f,i}> 0,\, l_i>0.
	\end{equation}
	We optimize problem \eqref{optim1} with respect to 
	$\left[c,\log\left(\bm{\sigma}_f\right),\log\left(\bm{\ell}\right)\right]$ (where the log is applied element-wise), which implicitly
	enforces the positive constraints and converts the original constrained problem to an unconstrained one. It is well-known that Eqn.~\eqref{optim1} is nonconvex with many local minima. Therefore, provided with its analytical derivatives \cite[Chapter~5]{williams2006gaussian}, we employ a derivative-based large-scale optimization algorithm, the truncated-Newton method \cite{di2017joint}, to locate the local optimal hyperparameters. Local convergence to a stationary point is guaranteed given Eqn.~\eqref{optim1} being  continuously differentiable and its derivative w.r.t. the hyperparameters being Lipschitz continuous \cite{dembo1982inexact}, in a neighborhood of a stationary point. 
	
	To summarize the overall workflow, given the prior mean/variance (i.e., $\bfm$ and $\bfK$) and assigned kernel, we maximize the log-likelihood by minimizing Eqn. \eqref{optim1}  to obtain the optimal hyperparameters, which are then used to update the prior mean/variance to obtain the posterior mean ($\bfm^*$) and variance ($\bfK^*$) using Eqn.~\ref{post_1}. For the optimization step, we initialize $c=0$,  $\log(\sigma_{f_i}) = 0$, and $l_i$ to be one pixel length. 
	
	\section{Numerical Results}
\begin{figure}[h]
	\centering
	\begin{tikzpicture}
		\node (A) at (0,0) {\includegraphics[width=0.16\textwidth]{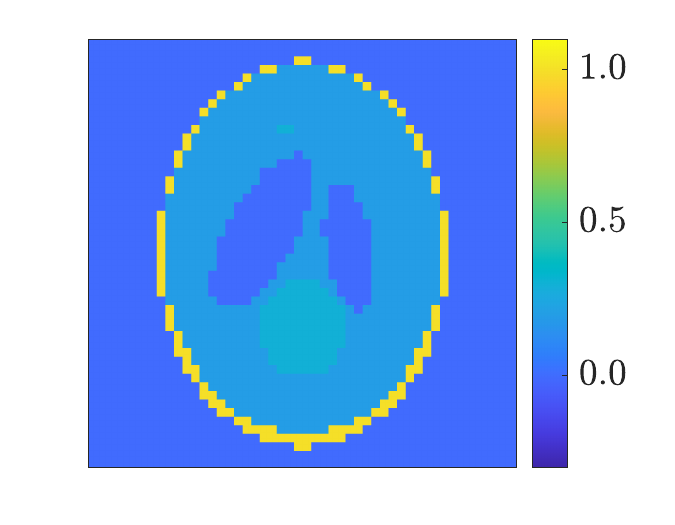}};
		\node[anchor=west, xshift=10pt] (B) at (A.east) {\includegraphics[width=0.165\textwidth]{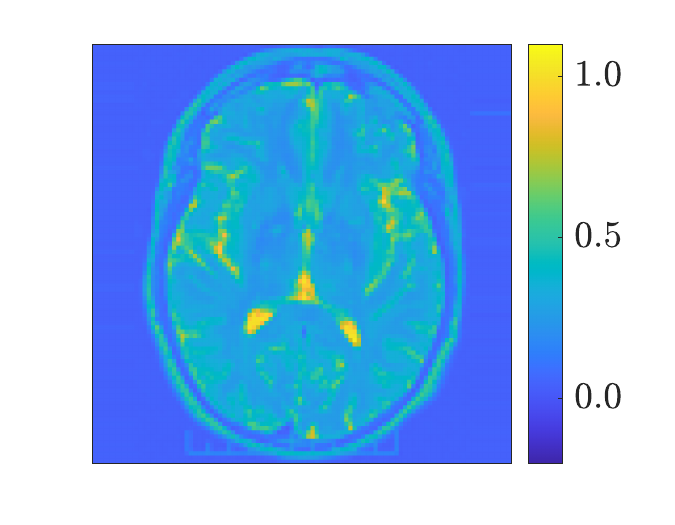}};
		\node[anchor=east,inner sep=0pt, xshift=5pt,yshift=-12pt] at (A.north west) {(a)};
		\node[anchor=east,inner sep=0pt, xshift=5pt,yshift=-12pt] at (B.north west) {(b)};
	\end{tikzpicture}
	\vspace{-1\baselineskip}
	\caption{(a) Shepp-Logan phantom, (b) Brain image}
	\label{fig:testObj}
	\vspace{5pt}
\end{figure}

	\vspace{-0.5em}
	We illustrate the performance of GP on the standard synthetic Shepp--Logan phantom (SLP) and a realistic brain image (see Figure~\ref{fig:testObj}). Both images are discretized into $n=100$ pixels along each dimension, and scanned with $N_\tau = 142$ beams for angles uniformly sampled between 0 and $2\pi$. We use the SLP phantom to test the GP method and establish its characteristics, while the brain image is used to further provide a comparative result with traditional optimization-based reconstruction methods. We consider the following two noise scenarios:
	\begin{description}
		\item[\namedlabel{itm:case1}{Case I}] Independent zero-mean Gaussian heteroscedastic noise with $\sigma_{\bfepsilon,t} = \alpha_t \text{RMS}(\tilde{\bfy})$ is added to each noise-free measurement $\tilde{y}_t$, 
		with $\alpha_t\sim \mathcal{U}(0.025, 0.100)$. The noise model for reconstruction correctly reflects the added heteroscedastic noise; \ie known $\alpha_t$-s for constructing $\bfSigma_\epsilon = \mathrm{diag}(\sigma^2_{\bfepsilon,1}, \sigma^2_{\bfepsilon,2}, \ldots, \sigma^2_{\bfepsilon,N_m})$.
		\item[\namedlabel{itm:case2}{Case II}] Same noise as \ref{itm:case1} is added to $\tilde{y}_t$, but assuming unknown $\alpha_t$-s. Instead, the noise is modeled to be zero-mean Gaussian with $\sigma_{\bfepsilon} = 0.05 \times \text{RMS}(\bfy)$ for all measurements and $\bfSigma_\epsilon = \mathrm{diag}(\sigma^2_{\bfepsilon}, \sigma^2_{\bfepsilon}, \ldots, \sigma^2_{\bfepsilon}) = \sigma_{\bfepsilon}^2 \bfI$.
\end{description}
The quality of the reconstruction is measured by the relative mean squared error as $\ds \text{E}_{\text{norm}} = \lvert\lvert \bff_{\bfr} - \bff^\dagger \rvert\rvert_F / \lvert\lvert \bff^\dagger \rvert\rvert_F,$ 
where $\bff^\dagger$ is the ground truth and $\bff_{\bfr}$ is the reconstruction returned by different algorithms (e.g., $\bfm^*$ from the GP method).

\begin{figure}[h]
	\centering
	\includegraphics[scale=1.8]{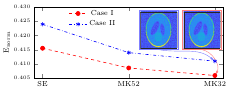}
	\caption{Comparison of GP-based reconstruction with different kernel types for the SLP object with $N_\theta=40$, MK32-based reconstructions shown as inset.}
	\label{new_SLP_comp_kernels}
	\vspace{5pt}
\end{figure}

We use a single kernel, \ie $N_k=1$, for the SLP object. Overall, the choice of kernel type appears to have little effect on the performance of the GP-based reconstruction as shown in \Cref{new_SLP_comp_kernels}, where we compare $\text{E}\textsubscript{norm}$ using three types of kernel---SE, MK32, and MK52---for Case I and II. Compared to \ref{itm:case2}, the reconstructions for all types of kernel are marginally better for \ref{itm:case1}, where the noise is modeled accurately. Since the MK32 kernel gives the best reconstruction, we will focus only on MK32 for further tests on SLP.

\begin{figure}
	\centering
	\includegraphics[width=0.80\columnwidth]{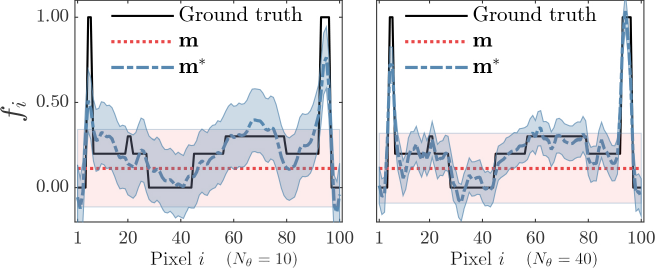}
	\caption{\ref{itm:case1} with MK32 kernel: Prior mean $\bfm$ and reconstructed mean $\bfm^\ast$ of pixels across the middle slice of SLP  for $N_\theta=10$ (left) and $N_\theta=40$ (right) angles. The shaded red and blue area represent the prior and posterior pixel-wise standard deviation around the respective means. 
	} 
	\label{slp_gpplot_case1}
\end{figure}

In \Cref{slp_gpplot_case1}, we compare the GP reconstruction (shown as the middle slice of the SLP image) using different numbers of projections as $N_\theta=10$ and $N_\theta=40$. As expected, more projections improve the reconstruction and, more importantly, reduces the associated uncertainty with the reconstruction. 

\begin{figure}
	\centering
	\includegraphics[width=0.45\columnwidth,trim={0 50 0 4},clip]{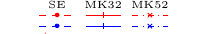}
	
	\includegraphics[width=0.9\columnwidth,trim={0 0 0 20},clip]{mri_loglik_err.pdf}
	
	\caption{Varying log-likelihood $\log p(\mathbf{y}|\mathbf{X})$ and E\textsubscript{norm} with increasing number of various component kernels $N_k$ for the brain image in (a) \ref{itm:case1} and (b) \ref{itm:case2}}
	\label{log_lik_and_rms_err_trend}
\end{figure}

Now we investigate the performance of composite kernels on a brain image with more complicated features that is scanned with $N_\theta=30$ angles. Figure~\ref{log_lik_and_rms_err_trend} shows the log-likelihood values and E\textsubscript{norm} using the optimized hyperparameters for different kernel types and their corresponding composites with various $N_k$. As expected, the log-likelihood plateaus as $N_k$ increases, given that the kernels are nested. Therefore, to avoid overfitting due to unnecessary component kernels, the optimal $N_k$ can be chosen greedily once the log-likelihood starts to stagnate to balance off the computational cost \cite{duvenaud2013structure}. In what follows, we show results a composite MK32 kernel with $N_k=2$ components.

\begin{figure}[htb]
	\centering
	\begin{tikzpicture}
		\node (A) at (0,0) {\includegraphics[width=0.34\columnwidth]{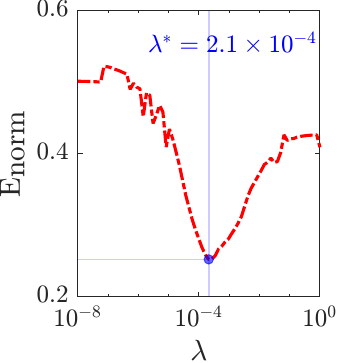}\label{mri_reg_wt_tv_recons}};
		\node[anchor=west,xshift=2em] (B) at (A.east) {\includegraphics[width=0.35\columnwidth]{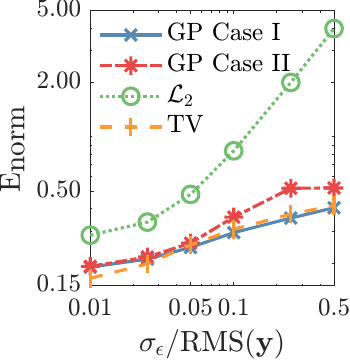}\label{mri_compare_recons_varying_snr}};
		\node[anchor=east,xshift=5pt,yshift=-10pt] (A1) at (A.north west) {(a)};
		\node[anchor=east,xshift=5pt,yshift=-10pt] (B1) at (B.north west) {(b)};
	\end{tikzpicture}
	\vspace{-0.5\baselineskip}
	\caption{(a) E\textsubscript{norm} for the TV-regularized reconstruction of the brain image with varying $\lambda$; (b) Comparison between GP-based reconstruction using $N_k = 2$ MK32 kernels and other approaches with varying expected noise.}
	\label{fig:combined_plots}
\end{figure}

\begin{figure}[htb]
	\centering
	\includegraphics[width=\columnwidth]{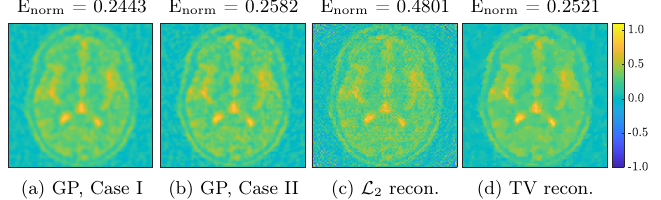}
	\vspace{-\baselineskip}
	\caption{Reconstruction comparison of the brain image.}
\label{mri_optim_recon_nonoise}
\end{figure}

\begin{figure}[h]
\centering
\includegraphics[width=0.70\columnwidth]{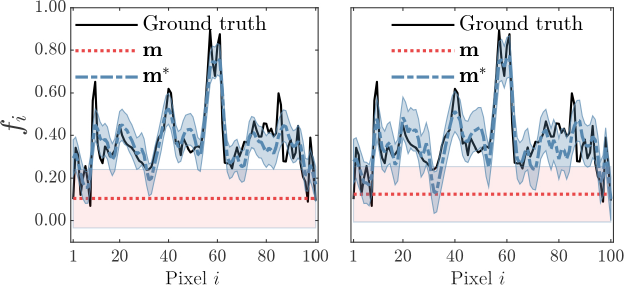}
\vspace{-0.5\baselineskip}
\caption{Prior mean $\bfm$ and posterior mean $\bfm^\ast$ using two MK32 kernels in \ref{itm:case1} (left) and \ref{itm:case2} (right). The shaded red and blue area represent the prior and posterior pixel-wise standard deviation around the respective means.}
\label{mri_gpplot}
\end{figure}

Next we compare the GP-based reconstruction with existing  techniques---$\mathcal{L}_2$ reconstruction (i.e., least squares) and the total variation (TV)- regularized $\mathcal{L}_2$ reconstruction \cite{huang2019calibrating}. We  explore  iterative  shrinkage-thresholding algorithms such as TwIST \cite{bioucas2007new} to efficiently solve the TV-regularized inverse problem. The optimal value $\lambda^\ast$, for the TV regularization parameter $\lambda$, is chosen as the one yielding the lowest E\textsubscript{norm} via a uniform grid search on $\lambda \in [10^{-8},10^{0}]$ (see Figure~\ref{fig:combined_plots}a). 
As a well-known challenge, the TV-based reconstruction quality is highly sensitive to the choice of $\lambda^\ast$, which makes it difficult for practical application in the absence of ground truth. In  contrast, the hyperparameters in the GP approach are chosen based on more principled approaches such as maximizing the likelihood of the observed data, such that the workflow is more systematic and robust. Figure~\ref{fig:combined_plots}(b) further compares performance at different noise levels. The noise added to the measurements is similar to \ref{itm:case1}, except for this study we let $\alpha_t \sim \mathcal{U}(\sigma_\epsilon - \Delta, \sigma_\epsilon + \Delta)$ where $\Delta = 0.5\sigma_\epsilon$, thus, as $\sigma_\epsilon$ increases the variance of $\alpha_t$-s also increase ultimately leading to higher levels of heteroscedasticity. For \ref{itm:case2} we assume $\alpha_t$-s are unknown, as before, and chose $\bfSigma_\epsilon = \sigma_\epsilon^2\bfI$. We emphasize that even with the optimally chosen parameter for TV-regularized reconstruction with access to the ground truth, the GP-based approach can still achieve comparable and even better reconstruction, especially at higher noise levels. Figure~\ref{mri_optim_recon_nonoise} shows the corresponding reconstructions, and Figure~\ref{mri_gpplot} shows the posterior mean and standard deviation for both Case I and II for the middle slice of the brain image, which again emphasizes the superb capability of the GP approach for providing a quantifiable measure of the reconstruction uncertainty.
\vspace{-\baselineskip}
\vspace{-1.2\baselineskip}
\section{Conclusion}
\vspace{-0.5em}
We explore GP modeling for tomography to not only provide a competitive reconstruction but also uncertainty quantification. We show the effect of noise characteristics and the number of measurements on the captured uncertainty. We also explore composite kernels that are capable of capturing different length scales in realistic objects and discuss a greedy technique for choosing the number of components. Numerical results show that the GP method greatly outperforms the $\mathcal{L}_2$ approach and achieves reconstruction comparable to TV-regularized solutions on the realistic brain image, without any dependence on ground truth. 


\bibliographystyle{IEEEbib}
\bibliography{strings,refs}

\begin{thebibliography}{10}

\bibitem{maire2014quantitative}
Eric Maire and Philip~John Withers,
\newblock ``Quantitative {X-ray} tomography,''
\newblock {\em International Materials Reviews}, vol. 59, no. 1, pp. 1--43,
  2014.

\bibitem{sharif2020comprehensive}
Muhammad~Irfan Sharif, Jian~Ping Li, Javeria Naz, and Iqra Rashid,
\newblock ``A comprehensive review on multi-organs tumor detection based on
  machine learning,''
\newblock {\em Pattern Recognition Letters}, vol. 131, pp. 30--37, 2020.

\bibitem{di2017joint}
Zichao~Wendy Di, Si~Chen, Young~Pyo Hong, Chris Jacobsen, Sven Leyffer, and
  Stefan~M Wild,
\newblock ``Joint reconstruction of {X}-ray fluorescence and transmission
  tomography,''
\newblock {\em Optics express}, vol. 25, no. 12, pp. 13107--13124, 2017.

\bibitem{antil2020bilevel}
Harbir Antil, Zichao~Wendy Di, and Ratna Khatri,
\newblock ``Bilevel optimization, deep learning and fractional {L}aplacian
  regularization with applications in tomography,''
\newblock {\em Inverse Problems}, vol. 36, no. 6, pp. 064001, 2020.

\bibitem{williams2006gaussian}
Carl~Edward Rasmussen and Christopher~KI Williams,
\newblock {\em Gaussian {P}rocesses for {M}achine {L}earning}, vol.~2,
\newblock MIT press Cambridge, MA, 2006.

\bibitem{he2011single}
He~He and Wan-Chi Siu,
\newblock ``Single image super-resolution using {G}aussian process
  regression,''
\newblock in {\em CVPR 2011}. IEEE, 2011, pp. 449--456.

\bibitem{wang2015single}
Haijun Wang, Xinbo Gao, Kaibing Zhang, and Jie Li,
\newblock ``Single-image super-resolution using active-sampling {Gaussian}
  process regression,''
\newblock {\em IEEE Transactions on Image Processing}, vol. 25, no. 2, pp.
  935--948, 2015.

\bibitem{li2013bayesian}
Dong Li, J~Svensson, H~Thomsen, F~Medina, A~Werner, and R~Wolf,
\newblock ``Bayesian soft {X-ray} tomography using non-stationary {Gaussian}
  processes,''
\newblock {\em Review of Scientific Instruments}, vol. 84, no. 8, pp. 083506,
  2013.

\bibitem{venkatakrishnan2016robust}
Singanallur~V Venkatakrishnan, Maryam Farmand, Young-Sang Yu, Hasti Majidi,
  Klaus van Benthem, Stefano Marchesini, David~A Shapiro, and Alexander
  Hexemer,
\newblock ``Robust {X}-ray phase ptycho-tomography,''
\newblock {\em IEEE Signal Processing Letters}, vol. 23, no. 7, pp. 944--948,
  2016.

\bibitem{purisha2019probabilistic}
Zenith Purisha, Carl Jidling, Niklas Wahlstr{\"o}m, Thomas~B Sch{\"o}n, and
  Simo S{\"a}rkk{\"a},
\newblock ``Probabilistic approach to limited-data computed tomography
  reconstruction,''
\newblock {\em Inverse Problems}, vol. 35, no. 10, pp. 105004, 2019.

\bibitem{beylkin1987discrete}
Gregory Beylkin,
\newblock ``Discrete {R}adon transform,''
\newblock {\em IEEE Transactions on Acoustics, Speech, and Signal Processing},
  vol. 35, no. 2, pp. 162--172, 1987.

\bibitem{ba2012composite}
Shan Ba and V~Roshan Joseph,
\newblock ``Composite {G}aussian process models for emulating expensive
  functions,''
\newblock {\em The Annals of Applied Statistics}, pp. 1838--1860, 2012.

\bibitem{rippa1999algorithm}
Shmuel Rippa,
\newblock ``An algorithm for selecting a good value for the parameter c in
  radial basis function interpolation,''
\newblock {\em Advances in Computational Mathematics}, vol. 11, no. 2, pp.
  193--210, 1999.

\bibitem{montegranario2014radial}
Hebert Montegranario and Jairo Espinosa,
\newblock ``Radial basis functions,''
\newblock in {\em Variational Regularization of 3{D} Data}, pp. 69--81.
  Springer, 2014.

\bibitem{dembo1982inexact}
Ron~S Dembo, Stanley~C Eisenstat, and Trond Steihaug,
\newblock ``Inexact {N}ewton methods,''
\newblock {\em SIAM Journal on Numerical analysis}, vol. 19, no. 2, pp.
  400--408, 1982.

\bibitem{duvenaud2013structure}
David Duvenaud, James Lloyd, Roger Grosse, Joshua Tenenbaum, and Ghahramani
  Zoubin,
\newblock ``Structure discovery in nonparametric regression through
  compositional kernel search,''
\newblock in {\em International Conference on Machine Learning}. Proceedings of
  Machine Learning Research, 2013, pp. 1166--1174.

\bibitem{huang2019calibrating}
Xiang Huang, Stefan~M Wild, and Zichao~Wendy Di,
\newblock ``Calibrating sensing drift in tomographic inversion,''
\newblock in {\em 2019 IEEE International Conference on Image Processing}.
  IEEE, 2019, pp. 1267--1271.

\bibitem{bioucas2007new}
Jos{\'e}~M Bioucas-Dias and M{\'a}rio~AT Figueiredo,
\newblock ``A new {TwIST}: Two-step iterative shrinkage/thresholding algorithms
  for image restoration,''
\newblock {\em IEEE Transactions on Image Processing}, vol. 16, no. 12, pp.
  2992--3004, 2007.

\end{thebibliography}

\end{document}